\documentstyle[epsfig,12pt]{article}
\def \bea{\begin{eqnarray}}
\def \beq{\begin{equation}}
\def \bo{B^0}

\def \eea{\end{eqnarray}}
\def \eeq{\end{equation}}

\def \ko{K^0}

\def \ob{\overline{B}^0}
\def \od{\overline{D}^0}

\def \pr{\parallel}
\def \s{\sqrt{2}}

\def \sx{\sqrt{6}}

\begin{document}
\begin{titlepage}

\large
\centerline {\bf Factorization vs. Flavor SU(3) in Charmless $B$ Decays
\footnote{ Enrico Fermi Institute preprint EFI 2000-41, hep-ph/0011183.
To be published in Proceedings of Beauty 2000, Kibbutz Maagan, Israel,
September 13--18, 2000, edited by S. Erhan, Y. Rozen, and P. E. Schlein,
Nucl.\ Inst.\ Meth. A, 2001.}}
\normalsize
 
\vskip 2.0cm
\centerline {Jonathan L. Rosner~\footnote{rosner@hep.uchicago.edu}}
\centerline {\it Enrico Fermi Institute and Department of Physics}
\centerline{\it University of Chicago, 5640 S. Ellis Avenue, Chicago, IL 60637}
\vskip 4.0cm
 
\centerline {\bf Abstract}
\vskip 1.0cm
Two types of predictions for charmless $B$ decays are compared.
One involves estimates based on factorization and models for form
factors, while the other involves the use of flavor SU(3), sometimes
with assumptions about the smallness of certain amplitudes.
After a comparison of some factorization
predictions with recent data, specific decays of $B$ mesons to two charmless
mesons are discussed.
\bigskip

\noindent
PACS Categories:  13.25.Hw, 14.409.Nd, 14.65.Fy, 12.39.Hg

\vfill
\end{titlepage}

\newpage
\section{Introduction}

The weak decays of hadrons are simpler when the weak current couples to a
lepton pair than when it couples to a hadron, which 
can re-interact with the rest of the system.
However, the effects of this re-interaction in some cases can be
neglected or evaluated, permitting the calculation of the decay rate
and individual helicity amplitudes.  In such cases one is employing the {\it
factorization hypothesis}.  An early version of this hypothesis \cite{BJ}
has recently been justified for certain classes of decays \cite{BBNS}.

We shall contrast applications of factorization,
which often require models for form
factors, with a more general approach based on flavor SU(3) in which data
are used to evaluate a set of reduced matrix elements but no assumptions are
made about form factors.
We review in Section 2 some successes of factorization in
final states containing a {\it heavy} meson.  Present data are compared
with early predictions \cite{JRFM}, and found
consistent with them.  The successful predictions involve 
{\it color-favored} cases, in which
weak currents couple to a quark pair ending up in the same meson.
By contrast, when the weak current produces a pair of quarks which end up in
{\it different} mesons ({\it color-suppressed} processes),
as well as for penguin diagrams, we shall question the applicability
of factorization.

We compare factorization and flavor SU(3) for $B$ meson decays
to charmless final states:  $B \to PP$ in Section 3 and $B \to PV$ in Section
4, where $P$ is a light pseudoscalar
meson ($\pi$, $K$, $\eta$, or $\eta'$) and $V$ is a light vector meson ($\rho$,
$\omega$, $K^*$, or $\phi$).
For $B \to PP$ we discuss the possible origin of the small $B \to \pi^+ \pi^-$
branching ratio.  We then mention recent applications of flavor SU(3) to
$B \to PP$ decays, and remark upon the large branching ratios for the decays
$B \to K \eta'$. For $B \to PV$ we compare an updated
flavor-SU(3) analysis \cite{GRPV} with others more
dependent upon factorization and explicit form factors
\cite{He,Hou,Xiao,Cheng,CYYY}.  We conclude in Section 5.

\section{Some successes of factorization}

\subsection{Semileptonic decays}

{\it Semileptonic} $b \to \ell \nu u$ or $b \to \ell \nu c$ decays
are tractable for several reasons.
(1) The weak current is a color singlet, so its effect
``factorizes.'' The leptons to which it couples can be treated in isolation
from the rest of the problem.  (2) When both the initial
and final hadrons are heavy, the decays are characterized by a single universal
(``Isgur-Wise'') form factor \cite{IW}.  (3) The form factors are
measurable via the effective mass distribution
of the lepton pair and the angular distributions of the decay products.

\subsection{Nonleptonic decays}

{\it Nonleptonic} decays involve
final-state interactions between the hadronic systems comprising
the two interacting weak currents.  The corresponding form factors are not
always measurable.   Nevertheless, in some cases these
nonleptonic decays can be treated by a factorization hypothesis.
 
\underline{Color-favored nonleptonic decays} of a $\bar b q$ meson
involve such subprocesses as
$\bar b \to \pi^+ \bar c$ or $\bar b \to \pi^+ \bar u$.  The coupling of
the weak current to the $\pi^+$ (or another light meson)
is described by a directly-measurable decay constant,
while the amplitude for the $\bar c$ or $\bar u$ quark to form a
hadron with the spectator quark $q$ is described by one or more measurable
form factors.  The interaction of the light meson with the rest of the
system is an effect of order $\Lambda_{QCD}/m_b$ and may be neglected for a
lowest-order estimate; a framework for calculating corrections has 
recently been established \cite{BBNS}.

\underline{Color-favored decays with current coupling to heavy mesons}
satisfy factorization (at least for $D_s$ and $D_s^*$ production), though
there is no corresponding theoretical justification.  In contrast to the
case in which the weak current produces a light meson, the heavy meson
produced by the weak current does not escape the interaction rapidly enough to
avoid significant interaction with the rest of the system \cite{BJ,BBNS}.

\underline{Color-suppressed or penguin amplitudes} are not of leading
order in $1/N_c$, where $N_c$ is the number of quark colors, or involve the
application of perturbative QCD under questionable circumstances.  The
corresponding factorization predictions for helicity amplitudes are not
obeyed, and the corresponding form factors are not directly measurable since
they involve an effective flavor-changing neutral current.

\subsection{Application to $D^{(*)} \ell \nu$ and $D^{(*)} \pi$ decays}

The decays $B \to \bar D^{(*)} \ell \nu$ and $B \to \bar D^{(*)} \pi$
involve a universal \cite{IW} form factor which is a function of the
dimensionless variable $z^2 \equiv (v - v')^2$, where $v$ and $v'$ are the
four-velocities of the initial and final heavy meson.  [One often speaks of the
variable $w = v \cdot v'$, related to $z$ by $z^2 = 2(1-w)$.]
Defining the four-momentum of the lepton pair or hadron to which the weak
current couples as $q$, and the dimensionless variable $y = q^2/m_B^2$, we can
write, adopting a single-pole \cite{JRFM} universal form factor:
\beq \label{eqn:ff}
z^2 = \frac{q^2 - (m_B - m_D)^2}{m_B m_D}~~,~~~
\frac{d \Gamma(B \to D^{(*)} \ell \nu)}
{d y} \sim \frac{{\rm kinem.~factor}}{[1 - (z^2/z_0^2)]^2}~~~,
\eeq
where $z_0$ is related to the slope
parameter $\rho^2$ (see, e.g., \cite{COV}) by $\rho^2 = 2/z_0^2$.

\begin{figure}
\centerline{\epsfysize = 5in \epsffile {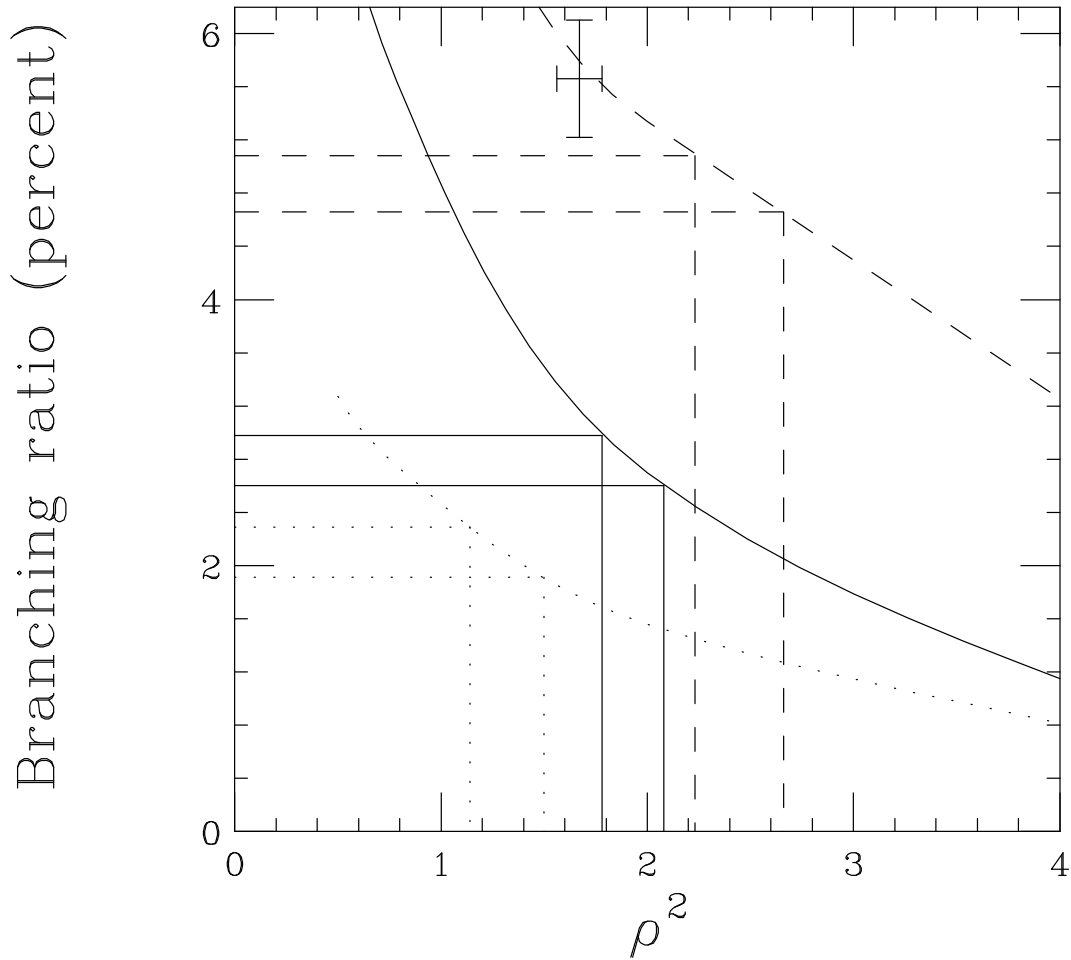}}
\caption{Factorization predictions for branching ratios (in percent) for some
$B$ decays based on the universal form factor (\ref{eqn:ff}), plotted
as functions of $\rho^2 = 2/z_0^2$.   Solid line: $10 {\cal B}
(B^0 \to D^{(*)-} \pi^+)$; dashed line: ${\cal B} (B^0 \to D^{*-} \ell^+
\nu)$; dotted line: ${\cal B} (B^0 \to D^- \ell^+ \nu)$.  Horizontal bands
denote $\pm 1 \sigma$ experimental limits; vertical bands denote corresponding
range of $\rho^2$.  Plotted point denotes value from Ref.\ \cite{CLEOVcb}.
\label{fig:D}}
\end{figure}

In Fig.\ \ref{fig:D} we plot some predictions \cite{JRFM} of rates for $B
\to \bar D^{(*)} \ell \nu$ and $B \to \bar D^{(*)} \pi$ decays as a
function of $\rho^2$.  Horizontal bands denote $\pm 1 \sigma$
error bars based on present averages \cite{PDG}.  The corresponding vertical
bands are consistent with a value of $\rho^2$ around 2 or $z_0$ around 1,
not far from the value $z_0 = 1.12 \pm 0.17$ found in Ref.\ \cite{JRFM}
(where the variable we now call $z$ was denoted by $w$).  Also shown is a
recent branching ratio, ${\cal B}(B^0 \to D^{*-} \ell^+ \nu) = (5.66 \pm 0.29
\pm 0.33)\%$, and a value $\rho^2 = 1.67 \pm 0.11$ reported by the CLEO
Collaboration \cite{CLEOVcb}.

\subsection{Current producing $D_s^{(*)}$ in $\ob$ decays}

In Table 1 we compare some factorization predictions \cite{JRFM} of light-meson
and $D_s^{(*)}$ production with experiment \cite{PDG,BaO}.  The $D_s^{(*)}$
predictions are as well obeyed as those for the light mesons.

An additional prediction involving heavy meson production by the weak current
\cite{JRFM} is that ${\cal B}(\bo \to D^{*+} D^{*-})/{\cal B}(\bo \to D^{*+}
D_s^-) = 0.13(f_D/f_{D_s})^2 \simeq 0.09$, where $f_D$ and $f_{D_s}$ are
the decay constants for the nonstrange and strange $D$ mesons.  The
experimental value for this ratio \cite{CLEODD} is $0.06^{+0.04}_{-0.03}$.

The decays of spinless particles to two
vector mesons are describable \cite{DDLR} by amplitudes
$A_0$ (longitudinal polarization), $A_\parallel$ (linear parallel polarization)
and $A_\perp$ (linear perpendicular polarization), normalized such that
$|A_0|^2 + |A_\parallel|^2 + |A_\perp|^2 = 1$.  Factorization predicts
$(|A_0|^2, |A_\parallel|^2, |A_\perp|^2) = (88,10,2)\%$ for $\ob \to D^{*+}
D_s^{*-}$ and (55,39,6)\% for $\ob \to D^{*+} \rho^-$.  Experimental values
are only quoted for $|A_0|^2$: $(87.8 \pm 3.4 \pm 3.0)\%$ for $\ob \to D^{*+}
D_s^{*-}$ \cite{CLEODD} and $(50.6 \pm 13.9 \pm 3.6)\%$ for $\ob \to D^{*+}
\rho^-$ \cite{dr}.  These agree with the predictions, as does the 
intermediate case of $\rho'(1418)$ production \cite{rhop}.

\subsection{Color-suppressed and penguin amplitudes}

The subprocess $\bar b \to \bar c c \bar s \to J/\psi \bar s$ is
is an example of color-suppression since the $\bar c c$ pair is not
automatically produced in a color singlet.  It is responsible for
the decays $B \to J/\psi K^{(*)}$.
The application of factorization to such decays is risky.
(1) There is no independent measurement of the $B \to K^{(*)}$ form factor,
which would involve a flavor-changing neutral current.  (2)
Factorization does not predict the helicity amplitudes properly in
$B \to J/\psi K^*$ \cite{facthel}.  (3) Final-state phases observed between
different helicity amplitudes for $B \to D^* \rho$ \cite{dr} and $B \to J/\psi
K^*$ \cite{psks} are not predictable by factorization, and may indicate the
importance of non-perturbative effects.  (4) QCD corrections to
color-suppressed amplitudes are important.  For example, the amplitude for $B^0
\to D^- \pi^+$ is purely color-favored, while that for $B^- \to \od \pi^-$
contains also a color-suppressed contribution which interferes constructively
with the color-favored amplitude.  This is in contrast to
charmed particle decays, where the color-suppressed and color-favored decays
interfere destructively.

Similar cautionary remarks apply to the use of factorization for penguin
amplitudes.  (1) Perturbative calculations of penguin contributions to
processes such as $B \to K \pi$, where they seem to be dominant, fall short of
actual measurements \cite{Ciu}.  One possible explanantion is the
presence of a $c \bar c$ loop with substantial enhancement from on-shell
states, equivalent to strong rescattering from such states as
$D_s \bar D$ to charmless meson pairs.  In this case,
penguin amplitudes could have different final-state phases from tree
amplitudes, enhancing the possibility of observing direct CP violation.
(2) Other hints that $c \bar c \to q \bar q$ rescattering may be important
include the suppression of the $B$ semileptonic
branching ratio with respect to theoretical expectations, the deficit
of charmed particles in $B$ decays, and the large rate for inclusive and
exclusive $\eta'$ production \cite{fs}.
 
\subsection{Further $B \to VV$ information}

For decays of a spinless particle to two vector mesons, the
amplitudes $A_0$ and $A_{\parallel}$ are linear combinations
of partial waves $\ell = 0,2$, while the amplitude $A_\perp$ corresponds
to $\ell = 1$.  In decays to CP eigenstates, the even- and odd-$\ell$
partial waves correspond to opposite CP-parities:  e.g., for $B_s \to J/\psi
\phi$, to even and odd CP, respectively.  The decay $B \to
J/\psi K^*$ is related to $B_s \to J/\psi \phi$ by flavor SU(3) \cite{DDLR},
so the helicity structures of the two decays should be the same.  In
Table 2 we compare CDF information on helicities for both decays \cite{psks}
with CLEO \cite{CLhel} and BaBar \cite{BaH} results on $B \to J/\psi K^*$.
In all cases the parity-odd fraction $|A_\perp|^2$ is small, indicating that
$B_s \to J/\psi \phi$ occurs mostly from the CP-even mixture of
$B_s$ and $\overline{B}_s$.  This will help in searching for lifetime
differences between the CP-even and CP-odd states \cite{DDLR,JRTASI}.
\medskip

A 1998 CLEO analysis \cite{dr} suggested non-zero relative final state
phases between partial waves in $B \to D^* \rho$.  Such final-state
phases are of interest in the more general context of final-state
interactions, which are usually thought to be small at a c.m. energy
of $m_b c^2$.  The existence of three partial waves (S,P,D) for such $B \to VV$
decays, as well as for final states with light mesons such as
$\bo \to \phi K^{*0}$, means that helicity analyses can detect the presence
of final-state interactions which could be relevant to the question of
rescattering and final-state interactions in decays such as $B \to PP$
\cite{fs}.

\section{Decays to two light pseudoscalars}

A flavor-SU(3) decomposition of the decays $B \to PP$, where $P$ denotes a
light pseudoscalar meson, is given in Table 3 \cite{etapx}.  Here $T,~C,~P$,
and $P_{EW}$ denote color-favored tree, color-suppressed tree, penguin, and
color-favored electroweak penguin (EWP) amplitudes, respectively.  We 
omit exchange, annihilation, and color-suppressed EWP amplitudes.  The
``singlet-penguin'' term $S'$ is needed whenever one final meson (such as
$\eta,\eta'$) has a flavor-singlet component.  These amplitudes
describe the decays in Table 4,
based on reports by CLEO \cite{CLEOkpi,CL2K}, BaBar \cite{BaO}, and
BELLE \cite{BEO} at the 2000 Osaka Conference,
and some earlier values \cite{GRVP}.  Remarks:

(1) The $K \pi$ rates should be dominated by penguin amplitudes.
Expanding to lowest order in the remaining amplitudes yields the sum
rule \cite{LSR}
\beq \label{eqn:LSR}
{\cal B}(K^+ \pi^-) + \frac{\tau^0}{\tau^+} {\cal B}(K^0 \pi^+) =
2 \left[ {\cal B}(K^0 \pi^0) + \frac{\tau^0}{\tau^+} {\cal B}(K^+ \pi^0)
\right]~~~,
\eeq
where $\tau^0/\tau^+ = 0.94 \pm 0.03$ is the ratio of $B^0$ and $B^+$
lifetimes.
The left-hand side is $32 \pm 4$, while the right-hand side is $57 \pm 11$.
The rates involving $\pi^0$ production may have
been overestimated experimentally; little could go
wrong with the sum rule unless penguin dominance turns out to have been a
very poor approximation.

(2) For now, complete penguin dominance of the amplitudes for $B \to K \pi$ is
as good as the sum rule (\ref{eqn:LSR}).  One doesn't see any compelling
pattern of the subsidiary (tree and EWP) amplitudes.

(3) The ratio $\Gamma(B^+ \to K^0 \pi^+)/2\Gamma(B^+ \to K^+ \pi^0)$ is
$0.69 \pm 0.22$.  Its value
can provide a constraint on the weak phase $\gamma = {\rm Arg}(V^*_{ub})$
\cite{NR}.

(4) The ratio $\Gamma(B^0 \to K^+ \pi^-)/\Gamma(B^+ \to K^0 \pi^+)$ is
$0.92 \pm 0.24$, compatible with 1. If it were less than 1, one could
establish an upper bound on $\sin^2 \gamma$ \cite{FM}.  For any value,
one can learn about $\gamma$ with the help of additional
information such as the CP-violating rate asymmetry in $\bo \to K^+ \pi^-$
and $\ob \to K^- \pi^+$ \cite{GR98}.

Several $B \to PP$ decays are amenable to a flavor-SU(3) treatment.

\subsection{Tree-penguin interference in $B^0 \to \pi^+ \pi^-$?}

We shall quote all rates in units of ($\bo$ branching ratio $\times
10^6$).  Thus, the average of $\bo \to \pi^+ \pi^-$ branching
ratios in Table 5 implies (updating \cite{GRVP})
\beq
|T|^2 + |P|^2 - 2 |TP| \cos \alpha \cos \delta = 5.6 \pm 1.3~~~,
\eeq
where $\alpha$ is a CKM phase and $\delta$ is a strong phase difference
between tree and penguin amplitudes.  From $B^+ \to \pi^+ \pi^0$ one infers
$|T+C|^2/2 = (4.6 \pm 2.0)(\tau^0/\tau^+)$
(see Tables 3 and 4), and with Re($C/T) = 0.1$
(cf.\ \cite{BBNS}) one estimates $|T| = 2.7 \pm 0.6$.  Meanwhile the penguin
amplitude can be estimated from $B^+ \to K^0 \pi^+$:  $|P'|^2 = (17.9 \pm
4.1)(\tau^0/\tau^+)$,
$|P'| = 4.1 \pm 0.5$, $|P| = \lambda|P'| = 0.9 \pm 0.1$, where
$\lambda \simeq 0.22$ is a parameter \cite{WP}
describing the hierarchy of CKM matrix elements.  Combining these results,
we find $\cos \alpha \cos \delta = 0.5 \pm 0.7$, less than $1 \sigma$
evidence for destructive interference.  Our conclusion is thus more
guarded than some based on factorization and explicit form factors
\cite{He,Hou}.

\subsection{Information on $K \eta'$ decays}

Given estimates of $|P'|$ and
$|P'_{EW}|$, and constructive $S'$--$P'$ interference, one needs a
modest singlet-penguin contribution $|S'| = (0.6 \pm 0.2)|P'|$ to account for
the large branching ratios for $B^+ \to K^+ \eta'$ and $\bo \to \ko \eta'$.
Its size may be a problem for explicit models (e.g.,
\cite{Xiao,Ali}), but not for flavor SU(3).
The $K \eta'$ rate is also large as a result of constructive interference
between nonstrange and strange quark contributions of $\eta'$ to the
ordinary penguin amplitude $P'$ \cite{HJLeta}.  Both
the singlet penguin and ordinary penguin contributions are much smaller
for the $K \eta$ final states \cite{etapx}, which are so far not observed.
 
\subsection{Measuring $\gamma$ with $K \pi$ decays}

The Fermilab Tevatron and the CERN Large Hadron Collider will produce
large numbers of $\pi^+ \pi^-$, $\pi^\pm K^\mp$, and $K^+ K^-$ pairs from
neutral $B$ mesons.
(1) The processes $B^0 \to K^+ K^-$ and $B_s \to \pi^+ \pi^-$ involve only
the spectator-quark amplitudes $E$ and $PA$, and thus should be suppressed.
They are related to one another by a flavor SU(3) ``U-spin'' reflection $s
\leftrightarrow d$ \cite{MGU}.
(2) The decays $B^0 \to \pi^+ \pi^-$ and $B_s \to K^+ K^-$ also are related to
each other by a U-spin reflection.  Time-dependent studies of both processes
allow one to distinguish strong and weak interaction information
and to measure the angle $\gamma$ \cite{DuFl}.  This appears
to be a promising method for Run II at the Fermilab Tevatron \cite{Wurt}. (3)
The decays of non-strange and strange neutral $B$ mesons to $K^\pm \pi^\mp$
provide another source of information on $\gamma$  when combined
with information on $B^+ \to \ko \pi^+$ \cite{bskpi}.  An error of $10^\circ$
on $\gamma$ seems feasible, and a relation between the CP-violating rate
asymmetries for the strange and nonstrange decays to $K^\pm \pi^\mp$ provides
a check of the flavor-SU(3) assumption.

\section{Decays to one pseudoscalar and one vector meson}

\subsection{Flavor SU(3) and $B \to PV$ decays}

For decays $B \to PV$ there are twice as many amplitudes as for $B \to PP$
since the spectator quark can end up either in the pseudoscalar meson or
the vector meson.  We label amplitudes with a subscript $P$ or $V$
corresponding to the meson containing the spectator, using small
letters to denote amplitudes corrected for EWP contributions.
Present data from CLEO \cite{CL2K,CLEOVP}, BaBar \cite{BaO}, and BELLE
\cite{BEO}) provide partial information on individual amplitudes.

One neglects special amplitudes associated with the flavor-singlet
components of the $\omega$ and $\phi$.  In contrast to the $\eta$ and $\eta'$,
these mesons couple to other matter only through connected quark
diagrams, respecting the Okubo--Zweig--Iizuka (OZI) rule.  The decay $B^+ \to
\pi^+ \omega$ is then dominated by $t_V$, while both $B \to \phi
K$ charge states are dominated by $p'_P$ (including a non-negligible
EWP contribution).  If the penguin amplitude involves an intermediate state
including only a quark-antiquark pair, Lipkin has argued that one must have
$p_V = -p_P$ \cite{GRVP,LipCC}.  In factorization models $p_V$ 
tends to be smaller in magnitude than this estimate.  This is a
key difference between the flavor-SU(3) and factorization approaches.

It is harder to learn $t_P$, which is expected to dominate $B^0 \to \rho^+
\pi^-$.  That decay must be distinguished via flavor-tagging
from $\ob \to \rho^+ \pi^-$, dominated by $t_V$.
Present data quote only the sum of the two modes \cite{CL2K,CLEOVP}.
Nonetheless, within wide errors it is possible to estimate $|t_P|^2$ and
$|t_V|^2$ separately.

\subsection{Comparison of data and predictions}

In Tables 5 and 6 we compare $B^+ \to PV$ and $B^0 \to PV$ data with
predictions of the flavor-SU(3) scheme \cite{GRVP} and factorization
models \cite{CYYY}.  The range of SU(3) predictions is given neglecting
tree-penguin interference, but rates can exceed or be less than the italicized 
values if $\gamma > 90^\circ$ or $\alpha < 90^\circ$ and strong final-state
phase differences are small.  In the opposite cases of $\gamma < 90^\circ$
or $\alpha > 90^\circ$ the rates can exceed or be less than the bold-faced
values.

The data must improve in accuracy to distinguish flavor-SU(3)
predictions from explicit models.  One needs to separate
$B^0 \to \rho^+ \pi^-$ from $\ob \to \rho^+ \pi^-$ in order to separate
$t_P$ from $t_V$ adequately.  Both approaches agree
on which ``signals'' should be real and which are statistical fluctuations.

There is hope of learning whether tree and penguin amplitudes
are interfering constructively (e.g., in $B^0 \to K^{*+} \pi^-$ and $B^+ \to
K^{*+} \eta$) or destructively in specific processes, but one cannot yet do
this reliably.  Let us see where present data stand on the first of these.

\subsection{Tree-penguin interference in $\bo \to K^{*+} \pi^-$}

The branching ratio for $\bo \to K^{*+} \pi^-$ quoted in Table 6 implies that
\beq \label{eqn:kpbr}
|p'_P|^2 + |t'_P|^2 - 2|t'_P p'_P| \cos \delta \cos \gamma =
22^{+9}_{-8} > 12~(90\% {\rm~c.l.})~~~,
\eeq
where we are using units of (b.r.$\times 10^6$).  At the same time, averaging
charged and neutral modes, ${\cal B}(B \to \phi K)$ implies
\beq
|p'_P - \frac{1}{6}p'_P|^2 = 6.2^{+2.0+0.7}_{-1.8-1.7}~~,~~~
|p'_P| = 3.0^{+0.5}_{-0.7}~~~,
\eeq
where the term $-(1/6)p'_P$ is an estimate of EWP effects \cite{GRVP,EWP}.
To estimate $|t'_P|$ we must use $B^+ \to (\rho^0,\omega) \pi^+$
and $\bo \to \rho^\mp \pi^\pm$ \cite{GRVP}, finding $|t_P| \le 4.5$,
$|t'_P| \le 1$.  Thus the inequality (\ref{eqn:kpbr}) weakly favors
constructive $t'_P$--$p'_P$ interference in $\bo \to K^{*+} \pi^-$.  For
$\cos \delta > 0$ this would require $\cos \gamma <0$, as has been claimed
in several factorization-based calculations \cite{He,Hou}.  Within
the more general flavor-SU(3) treatment a firm conclusion
requires many of the input branching ratios to be better measured.
 
\section{Conclusions}

\begin{itemize}

\item Na\"{\i}ve factorization works well for color-favored processes,
including some for which it is ``not proven.''

\item Be wary of ``factorization'' results for color-suppressed or penguin
amplitudes.  They actually contain considerable phenomenological input.

\item Flavor SU(3), supplemented with electroweak penguin calculations
and assumptions about rescattering (such as the neglect of
exchange and annihilation contributions) can lead to many useful relations,
e.g., between $B \to \pi \pi$ and 
$B_s \to K \bar K$, and between $B \to K \pi$ and $B_s \to K \pi$.

\item There is no problem in describing $B \to K \eta'$ decays as long as one
allows a singlet penguin amplitude, whose magnitude is probably a
nonperturbative effect.

\item $B \to PV$ decays are consistent with flavor SU(3) but more data will
be needed to test the predictions incisively and to compare them with those
of factorization and form-factor models.

\item Some interferences (e.g., in $B^0 \to \pi^+\pi^-$, $\bo \to K^{*+}\pi^-$,
and $B \to \eta K^*$) suggest $\gamma > 90^\circ$ or $\alpha < 90^\circ$ if
final-state phases are small, but the pattern is not yet compelling.

\end{itemize}

The wealth of forthcoming data from experiments at
Cornell, SLAC, KEK, and the hadron machines will make substantial progress on
these questions in the next few years.  At the same time we look forward to
more progress on proving the validity and limits of factorization.

\section*{Acknowledgments}

I thank Yoram Rozen, Peter Schlein, and
the other organizers of Beauty 2000 for
an enjoyable and informative conference, and
Prof.\ K. T. Mahanthappa at the University of Colorado
and colleagues at Cornell University, the University of Hawaii,
and the Technion for gracious hospitality
and fruitful interactions. I am indebted to 
Amol Dighe, Isard Dunietz, Michael Gronau, Harry J. Lipkin, Zumin Luo,
and Matthias Neubert for pleasant collaborations on these subjects.
This work was supported in part by the United
States Department of Energy through Grant No.\ DE FG02 90ER40560, and in part
by the U. S. -- Israel Binational Science Foundation through Grant No.\
98-00237.

\def \ajp#1#2#3{Am.\ J. Phys.\ {\bf#1} (#3) #2}
\def \apny#1#2#3{Ann.\ Phys.\ (N.Y.) {\bf#1} (#3) #2}
\def \app#1#2#3{Acta Phys.\ Polonica {\bf#1} (#3) #2}
\def \arnps#1#2#3{Ann.\ Rev.\ Nucl.\ Part.\ Sci.\ {\bf#1} (#3) #2}
\def \art{and references therein}
\def \cmts#1#2#3{Comments on Nucl.\ Part.\ Phys.\ {\bf#1} (#3) #2}
\def \cn{Collaboration}
\def \cp89{{\it CP Violation,} edited by C. Jarlskog (World Scientific,
Singapore, 1989)}
\def \efi{Enrico Fermi Institute Report No.\ }
\def \epjc#1#2#3{Eur.\ Phys.\ J. C {\bf#1} (#3) #2}
\def \f79{{\it Proceedings of the 1979 International Symposium on Lepton and
Photon Interactions at High Energies,} Fermilab, August 23-29, 1979, ed. by
T. B. W. Kirk and H. D. I. Abarbanel (Fermi National Accelerator Laboratory,
Batavia, IL, 1979}
\def \hb87{{\it Proceeding of the 1987 International Symposium on Lepton and
Photon Interactions at High Energies,} Hamburg, 1987, ed. by W. Bartel
and R. R\"uckl (Nucl.\ Phys.\ B, Proc.\ Suppl., vol.\ 3) (North-Holland,
Amsterdam, 1988)}
\def \ib{{\it ibid.}~}
\def \ibj#1#2#3{~{\bf#1} (#3) #2}
\def \ichep72{{\it Proceedings of the XVI International Conference on High
Energy Physics}, Chicago and Batavia, Illinois, Sept. 6 -- 13, 1972,
edited by J. D. Jackson, A. Roberts, and R. Donaldson (Fermilab, Batavia,
IL, 1972)}
\def \ijmpa#1#2#3{Int.\ J.\ Mod.\ Phys.\ A {\bf#1} (#3) #2}
\def \ite{{\it et al.}}
\def \jhep#1#2#3{JHEP {\bf#1} (#3) #2}
\def \jpb#1#2#3{J.\ Phys.\ B {\bf#1} (#3) #2}
\def \kaon{{\it Kaon Physics}, edited by J. L. Rosner and B. Winstein,
University of Chicago Press, 2000}
\def \lg{{\it Proceedings of the XIXth International Symposium on
Lepton and Photon Interactions,} Stanford, California, August 9--14 1999,
edited by J. Jaros and M. Peskin (World Scientific, Singapore, 2000)}
\def \lkl87{{\it Selected Topics in Electroweak Interactions} (Proceedings of
the Second Lake Louise Institute on New Frontiers in Particle Physics, 15 --
21 February, 1987), edited by J. M. Cameron \ite~(World Scientific, Singapore,
1987)}
\def \kdvs#1#2#3{{Kong.\ Danske Vid.\ Selsk., Matt-fys.\ Medd.} {\bf #1},
No.\ #2 (#3)}
\def \ky85{{\it Proceedings of the International Symposium on Lepton and
Photon Interactions at High Energy,} Kyoto, Aug.~19-24, 1985, edited by M.
Konuma and K. Takahashi (Kyoto Univ., Kyoto, 1985)}
\def \mpla#1#2#3{Mod.\ Phys.\ Lett.\ A {\bf#1} (#3) #2}
\def \nat#1#2#3{Nature {\bf#1} (#3) #2}
\def \nc#1#2#3{Nuovo Cim.\ {\bf#1} (#3) #2}
\def \nima#1#2#3{Nucl.\ Instr.\ Meth. A {\bf#1} (#3) #2}
\def \np#1#2#3{Nucl.\ Phys.\ {\bf#1} (#3) #2}
\def \npbps#1#2#3{Nucl.\ Phys.\ B Proc.\ Suppl.\ {\bf#1} (#3) #2}
\def \os{XXX International Conference on High Energy Physics, Osaka, Japan,
July 27 -- August 2, 2000}
\def \PDG{Particle Data Group, D. E. Groom \ite, \epjc{15}{1}{2000}}
\def \pisma#1#2#3#4{Pis'ma Zh.\ Eksp.\ Teor.\ Fiz.\ {\bf#1} (#3) #2 [JETP
Lett.\ {\bf#1} (#3) #4]}
\def \pl#1#2#3{Phys.\ Lett.\ {\bf#1} (#3) #2}
\def \pla#1#2#3{Phys.\ Lett.\ A {\bf#1} (#3) #2}
\def \plb#1#2#3{Phys.\ Lett.\ B {\bf#1} (#3) #2}
\def \pr#1#2#3{Phys.\ Rev.\ {\bf#1} (#3) #2}
\def \prc#1#2#3{Phys.\ Rev.\ C {\bf#1} (#3) #2}
\def \prd#1#2#3{Phys.\ Rev.\ D {\bf#1} (#3) #2}
\def \prl#1#2#3{Phys.\ Rev.\ Lett.\ {\bf#1} (#3) #2}
\def \prp#1#2#3{Phys.\ Rep.\ {\bf#1} (#3) #2}
\def \ptp#1#2#3{Prog.\ Theor.\ Phys.\ {\bf#1} (#3) #2}
\def \rmp#1#2#3{Rev.\ Mod.\ Phys.\ {\bf#1} (#3) #2}
\def \rp#1{~~~~~\ldots\ldots{\rm rp~}{#1}~~~~~}
\def \si90{25th International Conference on High Energy Physics, Singapore,
Aug. 2-8, 1990}
\def \slc87{{\it Proceedings of the Salt Lake City Meeting} (Division of
Particles and Fields, American Physical Society, Salt Lake City, Utah, 1987),
ed. by C. DeTar and J. S. Ball (World Scientific, Singapore, 1987)}
\def \slac89{{\it Proceedings of the XIVth International Symposium on
Lepton and Photon Interactions,} Stanford, California, 1989, edited by M.
Riordan (World Scientific, Singapore, 1990)}
\def \smass82{{\it Proceedings of the 1982 DPF Summer Study on Elementary
Particle Physics and Future Facilities}, Snowmass, Colorado, edited by R.
Donaldson, R. Gustafson, and F. Paige (World Scientific, Singapore, 1982)}
\def \smass90{{\it Research Directions for the Decade} (Proceedings of the
1990 Summer Study on High Energy Physics, June 25--July 13, Snowmass,
Colorado),
edited by E. L. Berger (World Scientific, Singapore, 1992)}
\def \tasi{{\it Testing the Standard Model} (Proceedings of the 1990
Theoretical Advanced Study Institute in Elementary Particle Physics, Boulder,
Colorado, 3--27 June, 1990), edited by M. Cveti\v{c} and P. Langacker
(World Scientific, Singapore, 1991)}
\def \yaf#1#2#3#4{Yad.\ Fiz.\ {\bf#1} (#3) #2 [Sov.\ J.\ Nucl.\ Phys.\
{\bf #1} (#3) #4]}
\def \zhetf#1#2#3#4#5#6{Zh.\ Eksp.\ Teor.\ Fiz.\ {\bf #1} (#3) #2 [Sov.\
Phys.\ - JETP {\bf #4} (#6) #5]}
\def \zpc#1#2#3{Zeit.\ Phys.\ C {\bf#1} (#3) #2}
\def \zpd#1#2#3{Zeit.\ Phys.\ D {\bf#1} (#3) #2}

\newpage

\begin{table}
\caption{Comparison of predictions for ratios of decay rates in
light-meson and $D_s^{(*)}$
production by the weak current in $\ob$ decays.  \label{tab:ds}}
\begin{center}
\begin{tabular}{|l|c|c|c|} \hline
Subprocess & $b \to c \bar u d$ & $b \to c \bar c s$ \\ \hline
Ratio      & $D^{*+} \pi^-/D^+ \pi^-$ & $D^{*+} D_s^-/D^+ D_s^-$ \\
Experiment &      $0.93 \pm 0.14$     &    $1.2 \pm 0.5$         \\
Prediction &            1             &          1               \\ \hline
Ratio      & $D^+ \rho^-/D^{*+}\pi^-$ & $D^+D_s^{*-}/D^{*+}D_s^-$ \\
Experiment &       $2.8 \pm 0.5$      &   $0.97 \pm 0.54$        \\
Prediction &           1.9            &          1               \\ \hline
Ratio    & $D^{*+}\rho^-/D^{*+}\pi^-$ & $D^{*+}D_s^{*-}/D^{*+}D_s^-$ \\
Experiment &       $3.4 \pm 0.8$      &    $2.2 \pm 0.6$         \\
Prediction &           2.2            &         2.6              \\ \hline
\end{tabular}
\end{center}
\end{table}

\begin{table}
\caption{Comparison of helicity amplitudes for $\bo \to J/\psi K^{*0}$ and
$B_s \to J/\psi \phi$. \label{tab:pshel}}
\begin{center}
\begin{tabular}{|c|c|c|} \hline
Amp.\ & CLEO ($\bo$) & CDF ($\bo$) \\ \hline
$|A_0|^2$ & $0.52 \pm 0.08$ & $0.59 \pm 0.06 \pm 0.01$ \\
$|A_\perp|^2$ & $0.16 \pm 0.09$ & $0.13^{+0.12}_{-0.09} \pm 0.06$ \\ \hline
Amp.\ & BaBar ($\bo$) & CDF ($B_s$) \\ \hline
$|A_0|^2$ & $0.60 \pm 0.06 \pm 0.04$ & $0.61 \pm 0.14 \pm 0.02$ \\
$|A_\perp|^2$ & $0.13 \pm 0.06 \pm 0.02$ & $0.23 \pm 0.19 \pm 0.04$ \\ \hline
\end{tabular}
\end{center}
\end{table}
 
\begin{table}
\caption{Flavor-SU(3) decomposition of some amplitudes for $B \to PP$, where
$P$ denotes a light pseudoscalar meson.  Unprimed amplitudes denote
strangeness-preserving decays; primed amplitudes denote strangeness-changing
decays.  \label{tab:PP}}
\begin{center}
\begin{tabular}{|c|c|} \hline
Mode & Amplitude \\ \hline
$\pi^+ \pi^-$ & $-(T + P)$ \\
$\pi^+ \pi^0$ & $-(T + C + P_{EW})/\s$ \\
$K^+ \pi^-$   & $-(T' + P')$ \\
$K^+ \pi^0$   & $-(T' + P' + C' + P'_{EW})/\s$ \\
$\ko \pi^+$   &        $P'$ \\
$\ko \pi^0$   & $(P'-C'-P'_{EW})/\s$ \\
$K^+ \eta'$   & $(3P'+4S'+T'+C'-\frac{1}{3}P'_{EW})/\sx$ \\
$\ko \eta'$   & $(3P'+4S'+C'-\frac{1}{3}P'_{EW})/\sx$ \\ \hline
\end{tabular}
\end{center}
\end{table}

\begin{table}
\caption{Branching ratios for some $B \to PP$ decays, in units of $10^{-6}$.
\label{tab:PPex}}
\begin{center}
\begin{tabular}{|c|c|c|c|c|} \hline
Mode & CLEO & BaBar & BELLE & Average \\ \hline
$\pi^+ \pi^-$ & $4.3^{+1.6}_{-1.4} \pm 0.5$ & $9.3 \pm 2.8$ & $6.3 \pm 4.0$ &
 $5.6 \pm 1.3$ \\
$\pi^+ \pi^0$ & $5.4 \pm 2.6$ & & $3.3 \pm 3.2$ & $4.6 \pm 2.0$ \\
$K^+ \pi^-$ & $17.2 \pm 2.7$ & $12.5 \pm 3.2$ & $17.4 \pm 5.9$ &
 $15.4 \pm 2.0$ \\
$\ko \pi^+$ & $18.2 \pm 4.6$ & & $16.6 \pm 9.1$ & $17.9 \pm 4.1$ \\
$K^+ \pi^0$ & $11.2 \pm 3.2$ & & $18.8 \pm 5.7$ & $13.0 \pm 2.8$ \\
$\ko \pi^0$ & $14.6 \pm 6.2$ & & $21.0 \pm 8.9$ & $16.7 \pm 5.1$ \\
$K^+ \eta'$ & $80 \pm 12$ & $62 \pm 20$ & & $75 \pm 10$ \\
$\ko \eta'$ & $89 \pm 19$ & & & $78 \pm 9$ (a) \\ \hline
\end{tabular}
\end{center}
\leftline{(a) Average for $K^+ \eta'$ and $K^0 \eta'$ modes.}
\end{table}

\begin{table}
\caption{Comparison of $B^+ \to PV$ data with predictions of flavor SU(3)
and factorization models.  Numbers denote predicted branching ratios in
units of $10^{-6}$.  See text for explanation of italicized and
bold entries.  \label{tab:bpvp}}
\begin{center}
\begin{tabular}{|c|c|c|c|} \hline
Mode & CLEO & Flavor SU(3) & Fact.\ models \\ \hline
$\rho^+ \pi^0$ & $< 43$ & {\it 4}--{\bf 15} & 10--13 \\
$\rho^0 \pi^+$ & $10.4^{+3.3}_{-3.4} \pm 2.1$ & {\bf 8}--{\it 17} & 9--13 \\
$\omega \pi^+$ & $11.3^{+3.3}_{-2.9} \pm 1.4$ & Input & 10--11 \\
$\rho^+ K^0$ & $< 48$ & 5--12 & \\
$\rho^0 K^+$ & $8.4^{+4.0}_{-3.4} \pm 1.8$ & {\it 1}--{\bf 2} & $\sim 1$ \\
$\omega K^+$ & $<7.9$ & {\it 2}--{\bf 4} & 1--2 \\
$\phi K^+$ & $6.4^{+2.5+0.5}_{-2.1-2.0}$ & Input & \\
$K^{*0} \pi^+$ & $7.6^{+3.5}_{-3.0} \pm 1.6$ & 5--12 & $\sim 4$ \\
$K^{*+} \pi^0$ & $< 31$ & {\bf 4}--{\it 7} & 4--6 \\
$K^{*+} \eta$ & $26.4^{+9.6}_{-8.2} \pm 3.3$ & {\bf 13}--{\it 22} & \\ \hline
\end{tabular}
\end{center}
\end{table}
 
\begin{table}
\caption{Comparison of $B^0 \to PV$ data with predictions of flavor SU(3)
and factorization models. Numbers denote predicted branching ratios in
units of $10^{-6}$.  See text for explanation of italicized and
bold entries.  \label{tab:bzvp}}
\begin{center}
\begin{tabular}{|c|c|c|c|} \hline
Mode & CLEO & Flavor SU(3) & Fact.\ models \\ \hline
$\rho^- \pi^+$ & $27.6^{+8.4}_{-7.4} \pm 4.2$ & {\bf 15}--{\it 32} & $\sim 14$
 \\
$\rho^+ \pi^-$ & (combined b.\ r.) & {\it 6}--{\bf 30} & $\sim 18$ \\
$\rho^- K^+$ & $16.0^{+7.6}_{-6.4} \pm 2.8$ & {\it 6}--{\bf 10} & 1--4 \\
$\rho^0 K^0$ & $< 27$ & 6--14 & \\
$\omega K^0$ & $10.0^{+5.4}_{-4.2} \pm 1.5$ & 2--3 & $0.4$ --2 \\
$\phi K^0$ & $5.9^{+4.0+1.1}_{-2.9-0.9}$ & Input & \\
$K^{*+} \pi^-$ & $22^{+8+4}_{-6-5}$ & {\bf 5}--{\it 10} & \\
$K^{*0} \pi^0$ & $< 3.6$ & 1--2 & 0.6--4 \\
$K^{*0} \eta$ & $13.8^{+5.5}_{-4.4} \pm 1.7$ & Input & \\ \hline
\end{tabular}
\end{center}
\end{table}
 
\newpage

\end{document}